\documentclass[twocolumn,
english,
superscriptaddress,
amsmath,
amssymb,
floats,
prb,
longbibliography]{revtex4-2}
\usepackage[latin9]{inputenc}
\usepackage{color}
\usepackage{mathtools}
\usepackage{amsmath}
\usepackage{graphicx}
\usepackage{multirow,booktabs}
\usepackage{array}
\usepackage{xcolor,colortbl}
\usepackage[normalem]{ulem}
\usepackage{lipsum}
\usepackage{soul}
\usepackage{dsfont}

\newcommand{\mbf}[1]{\mathbf{#1}}

\usepackage{array}
\newcommand{\PreserveBackslash}[1]{\let\temp=\\#1\let\\=\temp}
\newcolumntype{C}[1]{>{\PreserveBackslash\centering}p{#1}}
\newcolumntype{R}[1]{>{\PreserveBackslash\raggedleft}p{#1}}
\newcolumntype{L}[1]{>{\PreserveBackslash\raggedright}p{#1}}

\makeatletter

\usepackage{comment}

\usepackage[unicode=true,
 bookmarks=false,
 breaklinks=true,pdfborder={0 0 1},backref=false,colorlinks=true]
 {hyperref}
\hypersetup{pdfcreator={},pdfproducer={LaTeX with hyperref},linkcolor=blue,anchorcolor=blue,citecolor=blue,filecolor=red,menucolor=red,urlcolor=blue,pdfstartview=FitV,pdfhighlight=/I,hypertexnames=true}

\@ifundefined{textcolor}{}{%
 \definecolor{BLACK}{gray}{0}
 \definecolor{WHITE}{gray}{1}
 \definecolor{RED}{rgb}{1,0,0}
 \definecolor{GREEN}{rgb}{0,1,0}
 \definecolor{BLUE}{rgb}{0,0,1}
 \definecolor{CYAN}{cmyk}{1,0,0,0}
 \definecolor{MAGENTA}{cmyk}{0,1,0,0}
 \definecolor{YELLOW}{cmyk}{0,0,1,0}
}

\usepackage{epstopdf}
\usepackage{array}
\usepackage{mathrsfs}
\usepackage{dcolumn}
\usepackage{bm}

\usepackage{xcolor}

\newcolumntype{M}[1]{>{\centering\arraybackslash}m{#1}}
\newcolumntype{P}[1]{>{\centering\arraybackslash}p{#1}}

\newcolumntype{L}[1]{>{\raggedright\let\newline\\\arraybackslash\hspace{0pt}}m{#1}}
\newcolumntype{C}[1]{>{\centering\let\newline\\\arraybackslash\hspace{0pt}}m{#1}}
\newcolumntype{R}[1]{>{\raggedleft\let\newline\\\arraybackslash\hspace{0pt}}m{#1}}

\graphicspath{{../Inkscape/},{../Mathematica/},{./Figures/}}

\usepackage{tikz}
\usetikzlibrary{calc,intersections}
\usetikzlibrary{shadings}
\usepgflibrary{shadings}

\usepackage{babel}

\begin{document}

\title{Odd-parity magnetism in Fe-based superconductors with coplanar magnetic order}

\author{Reuel Dsouza}
\email{reuel.dsouza@nbi.ku.dk}
\affiliation{Niels Bohr Institute, University of Copenhagen, 2200 Copenhagen, Denmark}

\author{Andreas Kreisel}
\affiliation{Niels Bohr Institute, University of Copenhagen, 2200 Copenhagen, Denmark}

\author{Brian M. Andersen}
\affiliation{Niels Bohr Institute, University of Copenhagen, 2200 Copenhagen, Denmark}

\author{Daniel F. Agterberg}
\affiliation{Department of Physics, University of Wisconsin-Milwaukee, Milwaukee, Wisconsin 53201, USA} 

\author{Morten H. Christensen}
\email{mchriste@nbi.ku.dk}
\affiliation{Niels Bohr Institute, University of Copenhagen, 2200 Copenhagen, Denmark}

 \date{\today}
\begin{abstract}
Odd-parity magnetism constitutes an intriguing phase of matter which breaks inversion symmetry while preserving time-reversal symmetry. Here we demonstrate that the Fe-based superconductors exhibiting coplanar magnetic order realize an odd-parity magnetic state by combining low-energy modeling with density-functional theory. In the absence of spin-orbit coupling, the electronic spins are polarized along the $k_z$-direction and the splitting of the up and down states exhibits an $h$-wave form-factor. The magnitude of the splitting depends sensitively on specific parameters of the low-energy model, including specific out-of-plane hopping parameters and the Fermi energies of the hole- and electron-pockets. Interestingly, despite this state breaking inversion symmetry and exhibiting a finite out-of-plane Berry curvature and non-linear anomalous Hall effect, the Edelstein effect vanishes. Incorporating spin-orbit coupling tilts the momentum-space electronic spins into the ($k_x,k_y$)-plane and imparts finite in-plane components to the Edelstein response. Our findings highlight the Fe-based superconductors as platforms for exploring odd-parity magnetism both on its own and coexisting with unconventional superconductivity.
\end{abstract}
\maketitle

\section{Introduction}

Magnetism and superconductivity are hallmarks of quantum mechanics appearing at a macroscopic scale. The interplay of the two phenomena leads to a host of intriguing behaviors~\cite{Ran2019Nearly,Ran2019Extreme,Zhang2024Finite-momentum,Pientka2013Topological,Fulde1964Superconductivity,Larkin1964Nonuniform} and magnetism is believed to be crucial for the emergence of unconventional superconductivity~\cite{Scalapino2012A,Fernandes2022Iron,Monthoux2007Superconductivity,Romer2020}. Superconductivity is characterized by a condensate which can carry an internal angular momentum resulting in a non-trivial momentum-dependence. Magnetic analogues of such condensates -- where the spin-splitting has a non-trivial momentum-dependence -- have long been theorized~\cite{Wu2007Fermi} and the field has seen a rapid evolution in recent years~\cite{Hayami2019Momentum-Dependent,Hayami2020Spontaneous,Mazin2021Prediction,Smejkal2022Beyond,Mazin2022Editorial,Bhowal2022Ferroically,Hellenes2023P-wave}. This has led to the concepts of altermagnetism~\cite{Hayami2019Momentum-Dependent,Smejkal2022Emerging} -- with even parity spin-splitting -- and odd-parity magnetism~\cite{Hellenes2023P-wave} -- with odd parity spin-splitting. These ``unconventional magnets'' exhibit unusual properties such as large anomalous Hall effects, spin-polarized electrical currents, and a magneto-optical Kerr effect, that are usually associated with materials exhibiting strong spin-orbit coupling (SOC)~\cite{Smejkal2022Emerging,Smejkal2022Giant}. However, the interplay between the momentum dependencies of the superconducting and magnetic phases remains largely unexplored as materials exhibiting coexistence between unconventional forms of magnetism and superconductivity are currently lacking. While \emph{ab initio} methods and symmetry analyses have predicted a large number of candidate materials for hosting unconventional magnetism~\cite{Mazin2021Prediction,Smejkal2022Beyond,Hellenes2023P-wave,Guo2023Spin,Sodequist2024Two-dimensional,Yu2025Odd-parity,ChangheeLee2025}, altermagnetism has only been observed in a limited number of compounds~\cite{Krempasky2024Altermagnetic,Reimers2024Direct,Reichlova2024Observation,Regmi2025Altermagnetism,Candelora2026Discovery,Jiang2025A} and none currently feature a superconducting state. On the other hand, odd-parity magnetism has yet to be unambiguously experimentally observed though some recent works have found signatures of it~\cite{Yamada2025Odd,Song2025Electrical,Sears2025EuAuSb}. Here we show that the Fe-based superconductors exhibit odd-parity magnetism in a region of their phase diagram and thus constitute candidate materials for exploring the interplay between unconventional superconductivity and odd-parity magnetism.

\begin{figure}
    \centering
    \includegraphics[width=\columnwidth]{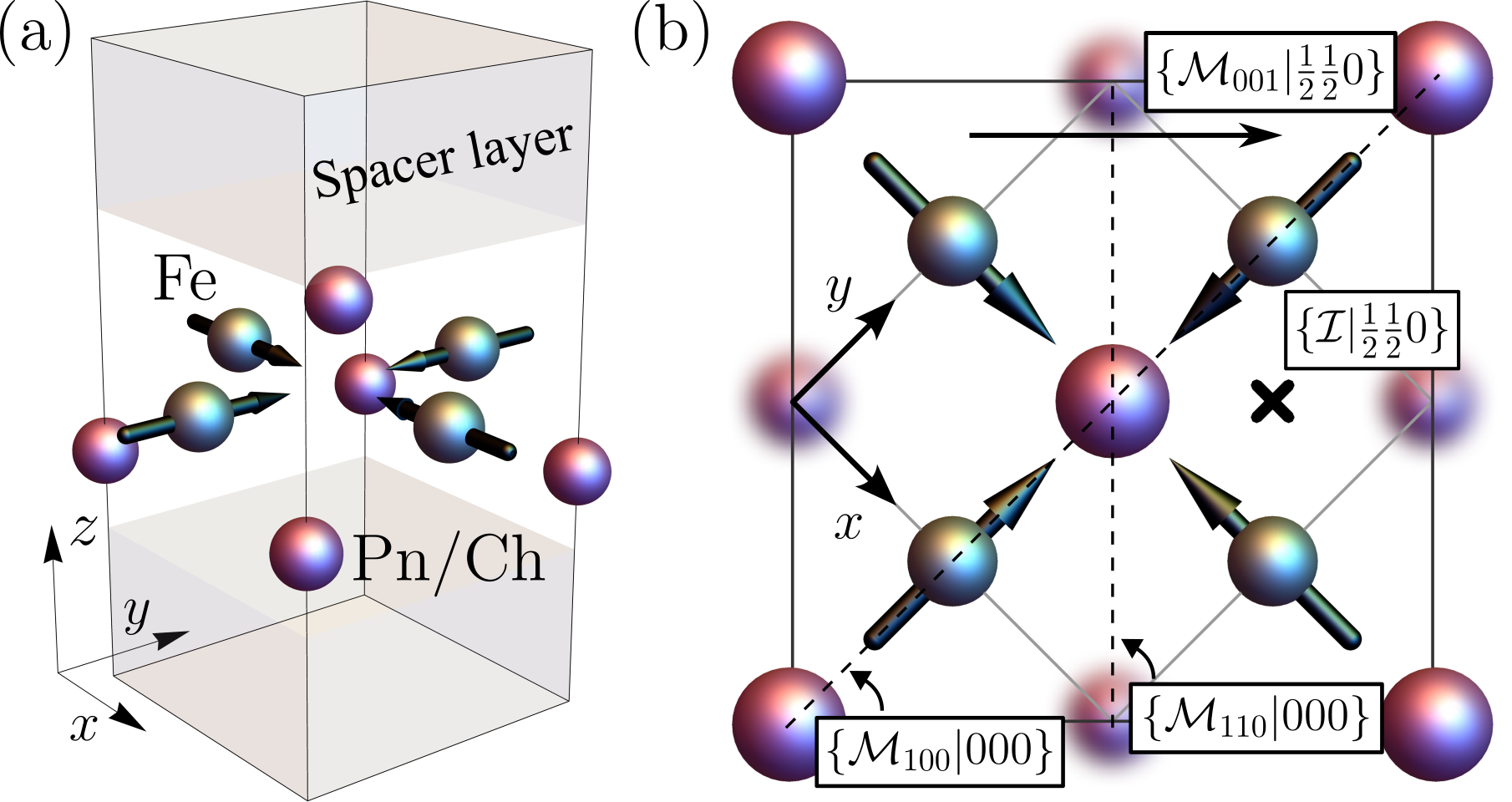}
    \caption{\label{fig:crystal_and_order} \textbf{Coplanar magnetic order in Fe-based superconductors.} (a) Crystal structure of a general Fe-based superconductor in the $P4/nmm$ space group with coplanar magnetic order and with Pn/Ch denoting a pnictogen or chalcogen atom. The spacer layer can contain, e.g., lanthanides or alkali metals, or be empty. (b) Top-down view of the coplanar magnetic order. The gray square denotes the original unit cell while the black square denotes the magnetic one. Note that the coordinate system is defined in the original cell and the origin is in the center. The mirror symmetries are denoted by dashed lines and the center of inversion is indicated by a cross. Note that here we only illustrate symmetries acting on real space.}
\end{figure}

The Fe-based superconductors are tetragonal systems consisting of Fe-layers with staggered pnictogen or chalcogen atoms above and below these layers~\cite{Fernandes2022Iron}. Some host an additional layer of donor atoms further from the Fe plane. The staggered nature of the pnictogen or chalcogen atoms leads to two distinct Fe atoms in the unit cell. The parent phase of these materials is metallic and their electronic structure are dominated by the Fe $d$-orbitals~\cite{Lu2008Electronic,Eschrig2009Tight-binding,Cvetkovic2013Space,Fernandes2017Low-energy,Kreisel2018On}. Most compounds feature a spin-density wave phase with moments on the Fe sites which is suppressed by doping resulting in superconductivity~\cite{Stewart2011Superconductivity} which, in some cases, coexist with the magnetic order~\cite{Stadel2022Multiple}. The two distinct Fe sites imply that there are two independent magnetic order parameters and consequently, the magnetic phase diagram displays three different phases~\cite{Allred2016Double-Q,Lorenzana2008Competing,Kang2015Interplay,Gastiasoro2015Competing,Christensen2015Spin,Christensen2017Role,Stadel2022Multiple}. One is the prolific collinear in-plane magnetic stripe phase, observed in most magnetic Fe-based superconductors~\cite{Lumsden2010Magnetism,Dai2015Antiferromagnetic}, while another is the collinear charge-spin density-wave phase in which the magnetic moments point out-of-plane and interfere destructively on every other site~\cite{Allred2016Double-Q,Avci2014Magnetically,Stadel2022Multiple}. Finally, there is a coplanar phase consisting of in-plane moments on every site oriented at right angles to each other~\cite{Meier2018Hedgehog,Stadel2022Multiple} (see Fig.~\ref{fig:crystal_and_order}) and this is the one relevant for odd-parity magnetism~\cite{Yu2025Odd-parity}. It has recently been observed in LaFeAs$_{1-x}$P$_x$O in a region of the phase diagram which also exhibits unconventional superconductivity~\cite{Stadel2022Multiple}, providing a materials platform in which to examine odd-parity magnetism driven superconducting phenomenology~\cite{Smidman2017Review} that includes spin-locked Cooper pairs~\cite{Sun2025Ising,Khodas2026Nonrelativistic-Ising} and non-trivial magnetoelectric responses~\cite{Shaffer2025Diode}.

Motivated by this finding, here we demonstrate that Fe-based superconductors in space group $P4/nmm$ ($\# 129$) with coplanar magnetic order are odd-parity magnets. Using a low-energy model, we show that the electronic bands generically exhibit spin-split bands with an $h$-wave form factor in the coplanar magnetic state. For coplanar order in the $xy$-plane, the momentum-space spin projection, $ S (\mbf{k}) $, is exclusively along the $k_z$-direction in the absence of SOC, and it exhibits nodes in the $(k_x, k_y)$, $(k_x, k_z)$, and $(k_y, k_z)$-planes and the diagonal $(k_x \pm k_y,k_z)$ planes. In the low-energy model, the spin splitting is controlled by the energy difference between states at $\Gamma$ and M as well as a combination of specific out-of-plane hopping parameters. Including SOC, the spin projection acquires components along $k_x$ and $k_y$ with $p$-wave character, while the $k_z$ component maintains its $h$-wave character. To connect with specific materials, we perform density functional theory (DFT) calculations for LaFeAsO and obtain a splitting of a few meV for a moment consistent with what is observed experimentally. This may be observable using, e.g., angle-resolved photoemission spectroscopy (ARPES), providing a realization of odd-parity magnetism in an unconventional superconductor.

This paper is organized as follows: In Sec.~\ref{sec:symmetries}, we discuss the coplanar magnetic order in more detail and argue - based on its spin space group symmetries - that it should exhibit odd-parity magnetism. In Sec.~\ref{sec:low_energy_model} we introduce an effective low-energy model for the Fe-based superconductors belonging to the $P4/nmm$ space group. Using this model, in Sec.~\ref{sec:odd_parity}, the coplanar magnetic phase is shown to result in an out-of-plane spin texture in momentum-space and the magnitude of the resulting spin-splitting is estimated based on the microscopic parameters of the model. The impact of SOC is discussed in Sec.~\ref{sec:SOC} and the spin texture in the collinear magnetic phase is briefly discussed in Sec.~\ref{sec:collinear}. Specific properties of the coplanar magnetic phase - the Edelstein effect and the intrinsic non-linear Hall effect - are evaluated in Secs.~\ref{sec:edelstein} and \ref{sec:non-linear-Hall}, respectively. In Sec.~\ref{sec:dft}, we present first-principles results for LaFeAsO in complete agreement with our low-energy modeling and we conclude in Sec.~\ref{sec:conclusions}. Technical details are presented in an Appendix.

\section{Symmetries of the coplanar magnetic order}\label{sec:symmetries}

Many Fe-based superconductors, including FeSe, LiFeAs, and LaFeAsO, crystallize in the nonsymmorphic $P4/nmm$ space group and are in that sense similar to CeNiAsO studied in Refs.~\cite{Hellenes2023P-wave,Yu2025Odd-parity}. CeNiAsO exhibits coplanar magnetic order with ordering vector $\mathbf{Q}_{\mathrm{X}}=(\pi,0,0)$ with the magnetic moments located on the Ce atoms occupying the $2c$ Wyckoff sites in the unit cell~\cite{Wu2019Incommensurate}. In contrast, the coplanar magnetic order in the Fe-based superconductors has ordering vector $\mathbf{Q}_{\mathrm{M}}=(\pi,\pi,0)$ and the moments sit on the Fe sites which occupy the $2b$ Wyckoff site. Such a coplanar state has recently been observed in LaFeAs$_{1-x}$P$_x$O near $x \approx 0.4$~\cite{Stadel2022Multiple} and its microscopic origin has been studied using both itinerant and localized approaches~\cite{Chubukov2008Magnetism,Lorenzana2008Competing,Si2008Strong,Kang2015Interplay,Gastiasoro2015Competing,Christensen2015Spin,Christensen2017Role,Yin2010Unified,Lv2010Orbital,Dai2012Magnetism}. Here, we will assume that such a magnetic state is condensed and focus on its consequences.

The coplanar magnetic state is shown in Fig.~\ref{fig:crystal_and_order}. In the language of spin-space groups this phase is invariant under the separate real- and spin-space mirror symmetries $\{ 2_{010} | \mathcal{M}_{100} \}$, $\{ 2_{\bar{1}10} | \mathcal{M}_{110} \}$, and $\{ 2_{\bar{1}10} | (\mathcal{M}_{001}|\tfrac{1}{2}\tfrac{1}{2}0 )\}$, where $\{\widehat{S}_g | \widehat{R}_g \}$ denotes operations on spin- ($\widehat{S}_g$) and real-space ($\widehat{R}_g$), respectively~\cite{Xiao2024Spin,Chen2024Enumeration,Jiang2024Enumeration}. The spatial part of the these symmetries are indicated in Fig.~\ref{fig:crystal_and_order}(b). Here, $\mathcal{M}_{100}$ denotes a mirror plane with normal along the $x$-direction and $2_{010}$ is a two-fold rotation around the $y$-axis. Taken together with the symmetry $\{ 2_{001} | 100 \}$, these imply that the coplanar phase exhibits odd-parity magnetism in momentum space, resulting in spin-split bands. In the absence of SOC, the momentum-space spin-polarization $ S (\mbf{k})=S_z(\mbf{k}) $ points along the $k_z$ axis although the above mirror symmetries enforce nodes in the spin-split bands along the $k_x=0$, $k_y=0$, and $k_z=0$ planes, as well as the diagonal $k_x = \pm k_y$ planes. This leads to a highly nodal $h$-wave form factor for the out-of-plane momentum-space spin texture to be discussed in detail below. Intriguingly, in the absence of SOC, $ S (\mbf{k}) $ is odd under all mirror symmetries, and the Edelstein effect vanishes identically, as we will show in Sec.~\ref{sec:edelstein}. This is contrast to, e.g., $p$-wave odd-parity magnets which exhibit a finite Edelstein response even in the absence of SOC~\cite{Yu2025Odd-parity}. On the other hand, the out-of-plane components of the nonlinear anomalous Hall effect~\cite{Sodemann2015Quantum, Du2021Nonlinear} are finite.

\begin{table}
    \centering
    \renewcommand{\arraystretch}{1.2} 
    \setlength{\tabcolsep}{6pt}       

    \begin{tabular}{cccccccc}
    \toprule
    \toprule
    $\varepsilon_{\Gamma}$ & $\varepsilon_1$ & $\varepsilon_3$ &
    $\tfrac{1}{2m_{\Gamma}}$ & $\tfrac{1}{2m_1}$ & $\tfrac{1}{2m_3}$ &
    $a_1$ & $a_3$ \\
    132 & -400 & -647 & -184 & 149 & 317 & 419 & -533 \\
    \midrule
    $b$ & $c$ & $v$ & $p_1$ & $p_2$ & $g_1$ & $t_{\Gamma,z}$ & $t_{M_1,z}$ \\
    56.5 & -62.3 & -243 & -40 & 10 & 100 & 7 & 7.1 \\
    \midrule
    $t_{M_3,z}$ & $\lambda_{\Gamma}$ & $\lambda_{\rm M}$ & & & & & \\
    7 & 25 & 25 & & & & & \\
    \bottomrule
    \bottomrule
    \end{tabular}

    \caption{\label{tab:parameters} Parameters used in the low-energy model (meV). All parameters except $g_1$, $t_{\{\Gamma,M_{1,3}\},z}$, and $\lambda_{\rm \Gamma,M}$ are taken from Ref.~\cite{Cvetkovic2013Space} and are based on fits to a tight-binding model for the Fe-based superconductors~\cite{Cvetkovic2009Multiband}. The dimensionless momenta $\mathbf{k}_\parallel$ and $k_z$ are expressed in units of $a^{-1}$ and $c^{-1}$, respectively.}
\end{table}

\section{Low-energy model}\label{sec:low_energy_model}

To describe the electronic impact of the odd-parity magnetic order we adopt an appropriate low-energy model for the Fe-based superconductors consisting of states at the $\Gamma$ and M points in the Brillouin zone. Using the irreducible representations (irreps) of the $P4/nmm$ space group at these points, Ref.~\cite{Cvetkovic2013Space} constructed a two-dimensional ($k_z=0$) $\mathbf{k}\cdot\mathbf{p}$-model including SOC, and classified the relevant magnetic order parameters.

Near the $\Gamma$ point, the relevant orbital states are the $d_{xz}$ and $d_{yz}$ orbitals from the two distinct Fe sites in the unit cell. In the basis
\begin{equation}
    \Psi_{\Gamma,\mbf{k}, \alpha}=\begin{pmatrix}
        \psi_{yz,\mbf{k},\alpha} \\ -\psi_{xz,\mbf{k},\alpha}
    \end{pmatrix}\,, \label{eq:Gamma_doublet}
\end{equation}
with $\alpha = \uparrow, \downarrow$ and the Hamiltonian is 
\begin{equation}
    \mathcal{H}_{\Gamma}=\sum_{\mbf{k}}\Psi^{\dagger}_{\Gamma,\mbf{k},\alpha}h_{\Gamma}(\mathbf{k})\Psi_{\Gamma,\mbf{k},\alpha}\,
\end{equation}
with
\begin{align}
    h_{\Gamma}(\mathbf{k}) &= \left[\varepsilon_{\Gamma} + \frac{k_x^2+k_y^2}{2m_{\Gamma}} + t_{\Gamma,z}(1-\cos k_z) \right] \tau^0 \nonumber \\ &
    + c(k_x^2 - k_y^2) \tau^1 + b k_x k_y \tau^3\,, \label{eq:ham_Gamma}
\end{align}
where $k_z$ is measured in units of the inverse $c$-axis lattice constant and $\boldsymbol{\tau}$ refers to Pauli matrices in orbital space. Near the M point there are two relevant orbital doublets, one consisting of the $d_{xz}$ and $d_{yz}$ orbitals of the two Fe sites, and one of the bonding and anti-bonding $d_{\pm xy}$ combinations from the two Fe sites. Respectively, these transform as the $M_1$ and $M_3$ space group irreps. In the bases
\begin{align}
    \Psi_{ M_1,\mbf{k},\alpha}&=\begin{pmatrix}
        \psi_{xz,\mbf{k}+\mbf{Q}_{\rm M},\alpha} \\ \psi_{yz,\mbf{k}+\mbf{Q}_{\rm M},\alpha}
    \end{pmatrix}\,, \nonumber \\
    \Psi_{M_3,\mbf{k},\alpha}&=\begin{pmatrix}
        \psi_{+xy,\mbf{k}+\mbf{Q}_{\rm M},\alpha} \\ \psi_{-xy,\mbf{k}+\mbf{Q}_{\rm M},\alpha}
    \end{pmatrix}\,,
\end{align}
the Hamiltonian matrices are
\begin{align}
    h_{M_1}(\mathbf{k}+\mbf{Q}_{\rm M}) &= \left[\varepsilon_{1} +  \frac{k_x^2+k_y^2}{2m_{1}} + t_{1,z}(1-\cos k_z) \right]\tau^0 \nonumber \\ &
    + a_1 k_x k_y \tau^3  + g_1 k_x k_y \sin k_z \tau^2\,, \label{eq:ham_M1}\\
    h_{M_3}(\mathbf{k}+\mbf{Q}_{\rm M}) &= \left[\varepsilon_{3} + \frac{k_x^2+k_y^2}{2m_{3}} + t_{3,z}(1-\cos k_z)\right]\tau^0 \nonumber \\ & + a_3 k_x k_y \tau^3\,.\label{eq:ham_M3}
\end{align}
We note that, for the states near M, $\mbf{k}$ is measured from $\mathrm{M}=(\pi,\pi,0)$. The term proportional to $g_1$ is important for obtaining the correct spin-splitting in the odd-parity magnetic state~\cite{Yu2025Odd-parity}. It originates from a hopping involving both in- and out-of-plane directions. These systems feature a weak out-of-plane dispersion and we have therefore extended the $k_z$-dependence to cover the range between $-\pi$ and $\pi$. Momentum-dependent terms couple the $d_{xz}/d_{yz}$ orbital states at M with the $d_{\pm xy}$ orbital states through the Hamiltonian
\begin{align}
    \mathcal{H}_{M_1 M_3} = \sum_{\mbf{k}} & i v_{+}(\mathbf{k}) \psi^{\dagger}_{+xy,\mbf{k}+\mbf{Q}_{\rm M},\alpha}\psi^{\phantom{\dagger}}_{xz,\mbf{k}+\mbf{Q}_{\rm M},\alpha} \nonumber \\  + & i v_{-}(\mathbf{k}) \psi^{\dagger}_{-xy,\mbf{k}+\mbf{Q}_{\rm M},\alpha}\psi^{\phantom{\dagger}}_{yz,\mbf{k}+\mbf{Q}_{\rm M},\alpha} + \text{H.c.}
\end{align}
where
\begin{align}
    v_{\pm}(\mbf{k}) &= v(\pm k_x + k_y) + p_1 (\pm k_x^3 + k_y^3) \nonumber \\ &+ p_2 k_x k_y (k_x \pm k_y)\,.
\end{align}

\begin{figure}[t]
    \centering
    \includegraphics[width=\columnwidth]{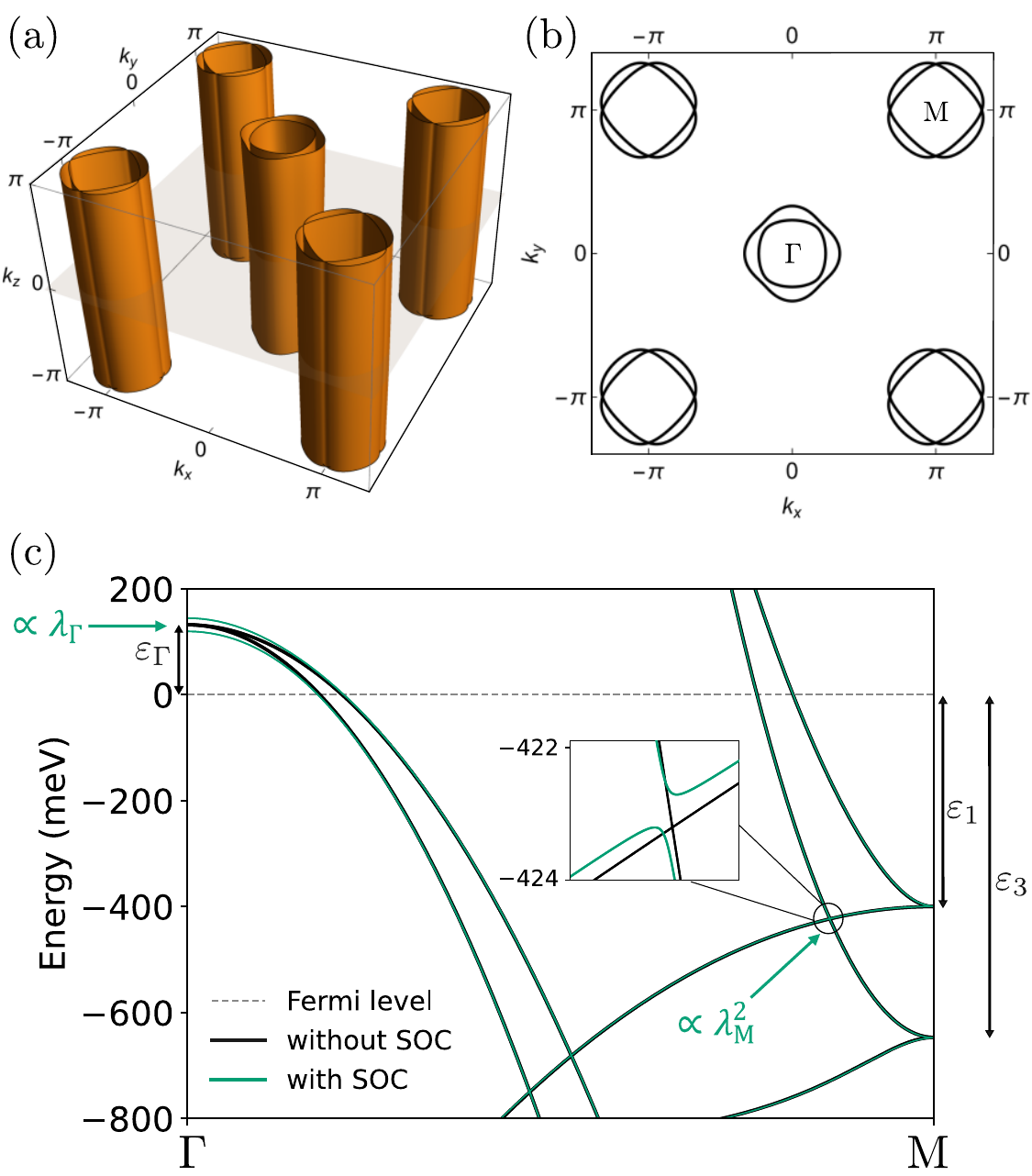}
    \caption{\label{fig:fs_disordered}\textbf{Normal state Fermi surface of low-energy model.} (a) Shows the Fermi surface of the low-energy model using the parameters listed in Table~\ref{tab:parameters} including the weak out-of-plane dispersion. (b) Shows a cut through the $k_z=0$ plane. (c) Electronic structure from $\Gamma$ to M with and without SOC.}
\end{figure}

The parameters used in the low-energy model are given in Table~\ref{tab:parameters}. The parameters controlling the in-plane dispersion are obtained from fitting the $\mbf{k}\cdot\mbf{p}$-model to the tight-binding model of Ref.~\cite{Cvetkovic2009Multiband}. In contrast, the parameters determining the out-of-plane dispersion are treated phenomenologically and are chosen to ensure a relatively weak out-of-plane dispersion. The Fermi surface of this low-energy model is shown in Fig.~\ref{fig:fs_disordered}(a) and a cut along the $k_z=0$ plane is shown in Fig.~\ref{fig:fs_disordered}(b). Here, both magnetic order and SOC are absent. The electronic structure from $\Gamma$ to M, both with and without SOC, is shown in Fig.~\ref{fig:fs_disordered}(c).

\begin{figure}
    \centering
    \includegraphics[width=1\columnwidth]{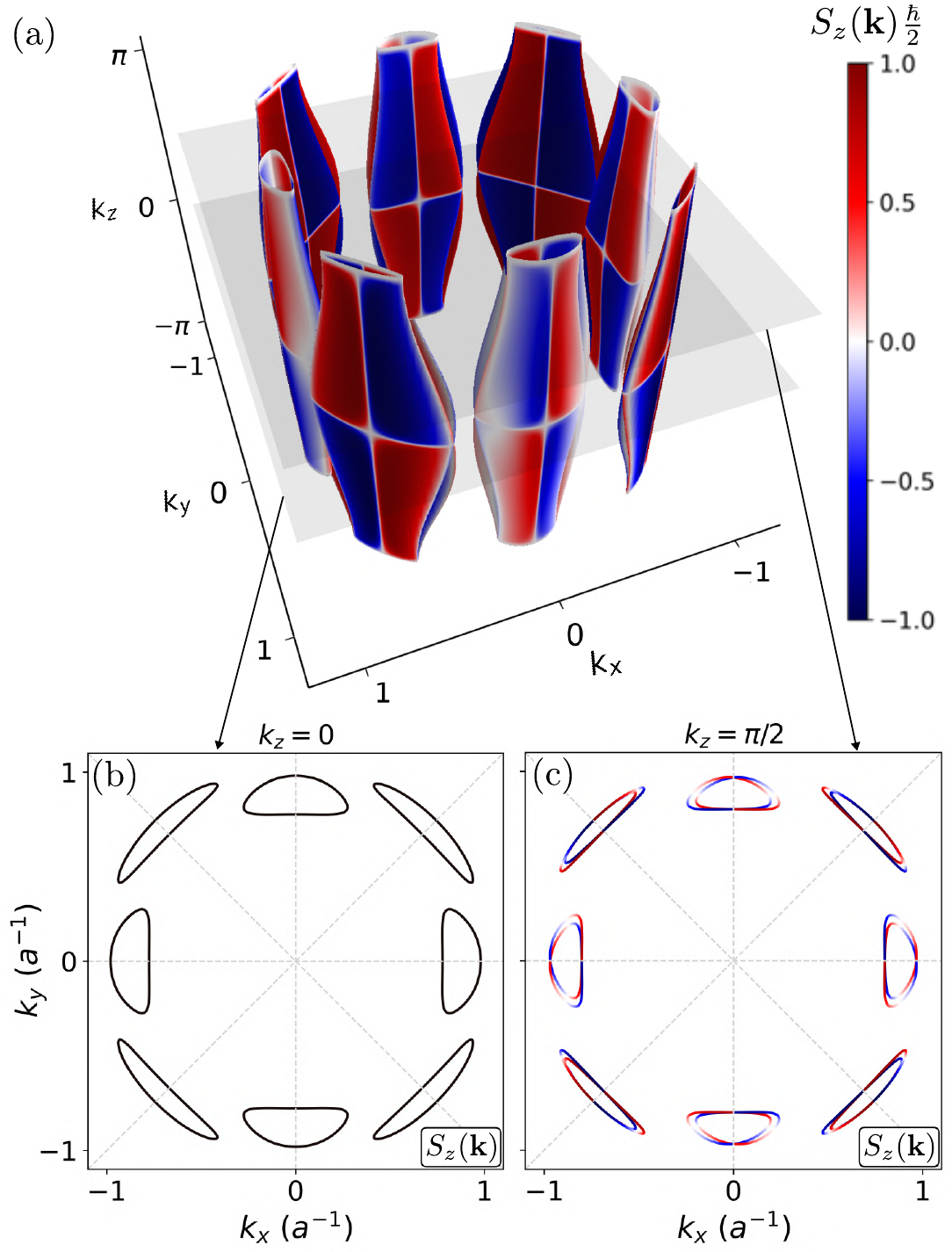}
    \caption{\label{fig:electronic_structure_no_SOC} \textbf{Momentum-space spin texture in the coplanar phase without SOC.} (a) Fermi surface in the coplanar magnetic phase colored according to $ S_z (\mbf{k})$. (b) and (c) show cuts of the Fermi surface at $k_z=0$ and $k_z=\pi/2$, respectively, showing how the bands split away from the $k_z=0$ plane. In (b) the Fermi surface is highlighted in black to make it visible. Here, the magnetic order parameter $\Delta=60$ meV.}
\end{figure}

\section{Odd-parity magnetism}\label{sec:odd_parity}

As discussed above, the magnetic ordering vector in the coplanar phase is $\mathbf{Q}_{\mathrm{M}}=(\pi,\pi,0)$~\cite{Stadel2022Multiple}. In the low-energy model, the coplanar magnetic order is diagonal in orbital space~\cite{Cvetkovic2013Space} thus coupling the $d_{xz}$ orbital states near $\Gamma-\mathrm{Z}$ to $d_{xz}$ states near $\mathrm{M}-\mathrm{A}$ and similarly for $d_{yz}$. This leads to the Hamiltonian
\begin{align}
    \mathcal{H}_{\rm mag} &= \Delta \sum_{\mathbf{k}}\sum_{\alpha\beta} \Big[ \psi^{\dagger}_{xz,\mbf{k},\alpha}\sigma^y_{\alpha\beta}\psi_{xz,\mbf{k}+\mbf{Q}_{\mathrm{M}},\beta} \nonumber \\ &+ 
     \psi^{\dagger}_{yz,\mbf{k},\alpha}\sigma^x_{\alpha\beta}\psi_{yz,\mbf{k}+\mbf{Q}_{\mathrm{M}},\beta} \Big] +\text{H.c}\,, \label{eq:ham_mag}
\end{align}
where $\Delta$ denotes the coplanar magnetic order parameter. Note that, for technical reasons, the spin coordinate-system lies along the Fe-Fe bonds~\cite{Cvetkovic2013Space}. This order breaks inversion symmetry but preserves a generalized form of time-reversal symmetry as well as a two-fold rotation in spin-space coupled with a translation in real space which leads to odd-parity magnetism. The resulting Fermi surface and spin-polarization is shown in Fig.~\ref{fig:electronic_structure_no_SOC}(a) exhibiting the expected $h$-wave structure. In the low-energy model, the spin splitting can be expressed as (see the Appendix)
\begin{equation}
    \Delta_E \sim |\Delta|^2\frac{ 2 a_1  c g_1}{(\varepsilon_\Gamma - \varepsilon_1)^2} f(\mathbf k) (k_x^2 - k_y^2) k_x k_y \sin k_z, \label{eq:spin_splitting}
\end{equation}
where $f(\mathbf{k})$ is a non-vanishing function of $\mathbf{k}$, the specific form of which is given in the appendix Eq.~\eqref{eq:f_k_function}. Note that $\epsilon_1$ is negative as it describes the energy of the electronic states at M. As predicted in Ref.~\cite{Yu2025Odd-parity}, the splitting vanishes if $g_1$ is zero. More importantly, the splitting is inversely proportional to the Fermi energy of the states at $\Gamma$ ($\varepsilon_{\Gamma}$) and M, ($\varepsilon_1$). Note that the fact that $\varepsilon_3$ is not referenced here is due to the magnetic order, Eq.~\eqref{eq:ham_mag}, in the low-energy model being diagonal in orbital space. From Eq.~\eqref{eq:spin_splitting}, the spin splitting is larger for materials with smaller Fermi energies and larger values of $g_1$, which scales with the magnitude of the out-of-plane hopping parameters. This is consistent with the fact that DFT calculations predict a larger spin splitting in FeSe than in LaFeAsO, as we show in Sec.~\ref{sec:dft}. FeSe has the smallest Fermi energy of all the Fe-based superconductors and it lacks a spacer-layer which increases the inter-layer coupling, thus allowing $g_1$ to be sizable. However, no coplanar magnetic order has been observed in FeSe.
%

\subsection{Impact of spin-orbit coupling}\label{sec:SOC}

The Fermi energy in the Fe-based superconductors is generally quite small, ranging from 10-20 meV in FeSe~\cite{Coldea2018The} to a few hundred meV for, e.g., LaFePO~\cite{Lu2008Electronic}. In comparison, the SOC is estimated to lie in the range between 10 and 25 meV~\cite{Borisenko2016Direct}, significantly impacting the electronic states near the Fermi level. To see how this influences the predicted odd-parity magnetic phase, we incorporate atomic SOC through the terms~\cite{Cvetkovic2013Space}
\begin{align}
  \mathcal{H}^{\rm SOC}_{\Gamma} &= \frac{i}{2}\lambda_{\Gamma}\sum_{\mbf{k}}\psi^{\dagger}_{xz,\mbf{k},\alpha} \sigma^z_{\alpha\beta} \psi_{yz,\mbf{k},\beta} + \text{H.c.}  \\
    \mathcal{H}^{\rm SOC}_{\rm M} &= \frac{i}{2}\lambda_{\rm M}\sum_{\mbf{k}}\Big( \psi^{\dagger}_{xz,\mbf{k}+\mbf{Q}_{\mathrm{M}},\alpha}\sigma^x_{\alpha\beta} \psi_{+xy,\mbf{k}+\mbf{Q}_{\mathrm{M}},\beta} \nonumber \\ & + \psi^{\dagger}_{-xy,\mbf{k}+\mbf{Q}_{\mathrm{M}},\alpha}\sigma^y_{\alpha\beta} \psi_{yz,\mbf{k}+\mbf{Q}_{\mathrm{M}},\beta} \Big) + \text{H.c.}\,.
\end{align}
Here, $\lambda_{\Gamma}$ is directly related to the energy splitting of the states at $\Gamma$ that are degenerate without SOC. In contrast, $\lambda_{\rm M}$ couples states that are non-degenerate, and its impact on the electronic structure is less direct, leading to hybridization between states away from the M point, which scales quadratically with $\lambda_{\rm M}$ as seen in Fig.~\ref{fig:fs_disordered}(c).

\begin{figure}
    \centering
    \includegraphics[width=.76\linewidth]{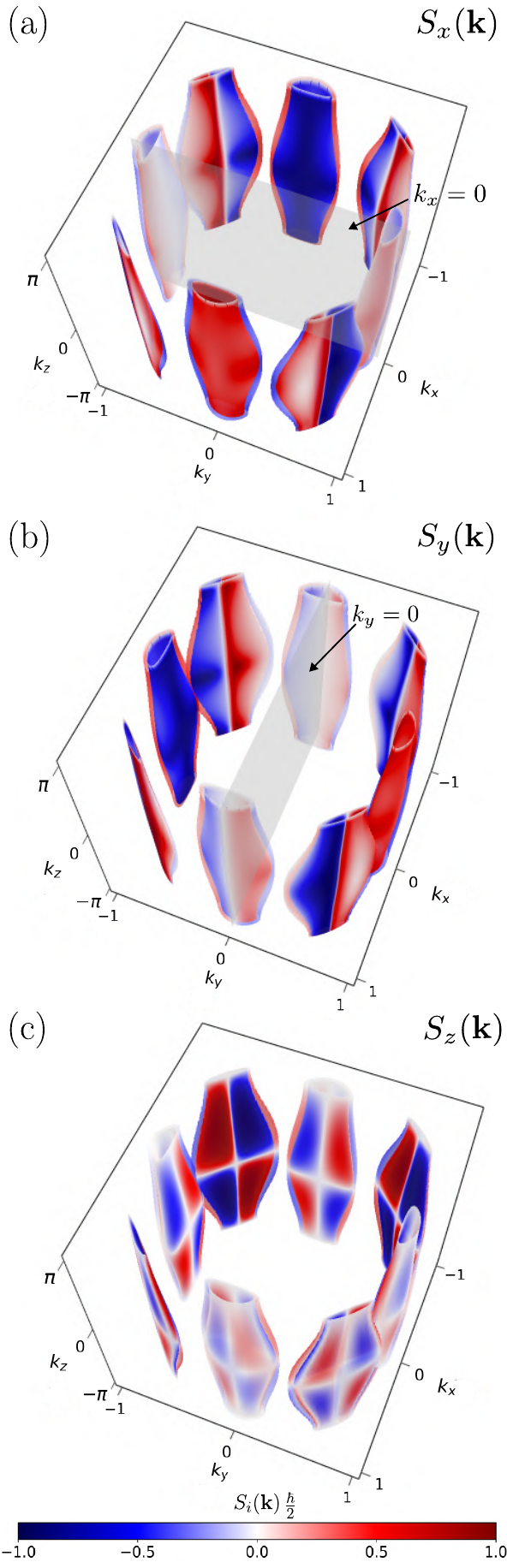}
    \caption{\label{fig:SOC_impact} \textbf{Impact of spin-orbit coupling.} Fermi surfaces showing (a) $ S_x(\mbf{k}) $, (b) $ S_y(\mbf{k}) $, and (c) $ S_z(\mbf{k}) $. The gray planes in (a) and (b) denote the mirror planes with respect to which the spin projection is antisymmetric; $ S_x(\mbf{k}) $ transforms as $p_x$, $ S_y(\mbf{k})$ as $p_y$, and $S_z(\mbf{k})$ maintains its $h$-wave structure. Note that the mirror planes in (c) are not highlighted. Here, $\Delta=60$ meV.}
\end{figure}
A finite SOC breaks the independence of the real- and spin-space transformations which underlies the presence of nodal planes in the electronic structure, although we note that the real-space spin-components are still confined to lie in the Fe-plane~\cite{Cvetkovic2013Space}. Consequently, finite values of $\lambda_{\Gamma}$ and $\lambda_{\rm M}$ renders the electronic structure fully non-degenerate, except for accidental degeneracies. This forces the momentum-space spin-polarization to acquire in-plane components and while the out-of-plane component of the spin still exhibits $h$-wave symmetry, the in-plane components do not, as shown in Fig.~\ref{fig:SOC_impact}. Instead, these exhibit $p_x$- and $p_y$-wave form-factors. This can be understood from the Rashba-like SOC terms induced by the coplanar order~\cite{Christensen2019Intertwined}. An example of one such term is
\begin{align}
    \mathcal{H}_{\lambda_{\Delta}} &= \lambda_{\Delta}\sum_{\mbf{k}} \left[ k_x (\sigma^x_{\alpha\beta} + \sigma^y_{\alpha\beta}) + k_y (\sigma^y_{\alpha\beta} - \sigma^x_{\alpha\beta}) \right]  \nonumber \\ & \times\left(\psi^{\dagger}_{xz,\mbf{k},\alpha}\psi_{xz,\mbf{k},\beta} + \psi^{\dagger}_{yz,\mbf{k},\alpha}\psi_{yz,\mbf{k},\beta} \right)\,,\label{eq:vestigial_SOC}
\end{align}
where $\lambda_{\Delta}$ depends linearly on $\lambda_{\Gamma}$ and quadratically on $\Delta$, the coplanar magnetic order parameter~\cite{Christensen2019Intertwined}, implying that this is a small effect. Henceforth, we will rotate the spin-coordinate system to align with the momentum space one. This implies $ \left[S_x(\mbf{k}) +S_y(\mbf{k})\right]/\sqrt{2} \rightarrow S_x(\mbf{k})$ and $ \left[ S_y(\mbf{k}) - S_x(\mbf{k}) \right]/\sqrt{2} \rightarrow S_y (\mbf{k})$. This makes the $p$-wave nature in Figs.~\ref{fig:SOC_impact}(a) and (b) explicit. We note that there are two Fermi surface sheets in Figs.~\ref{fig:SOC_impact}(a)--(c), the outer one is made semi-transparent.

The impact of SOC is most manifest near the original mirror planes of the $h$-wave structure. It ensures that the momentum-space spin direction is defined everywhere on the Fermi surfaces. This can be seen from Fig.~\ref{fig:SOC_impact}. Here, it is important to note that what looks like diagonal nodal planes in Figs.~\ref{fig:SOC_impact}(a) and (b) do not lie exactly along the diagonals but are offset compared to the diagonal mirror planes of Fig.~\ref{fig:SOC_impact}(c). 



\subsection{Aside: Spin-splitting in the collinear phase}\label{sec:collinear}

\begin{figure}[t]
        \centering
        \includegraphics[width=.8\linewidth]{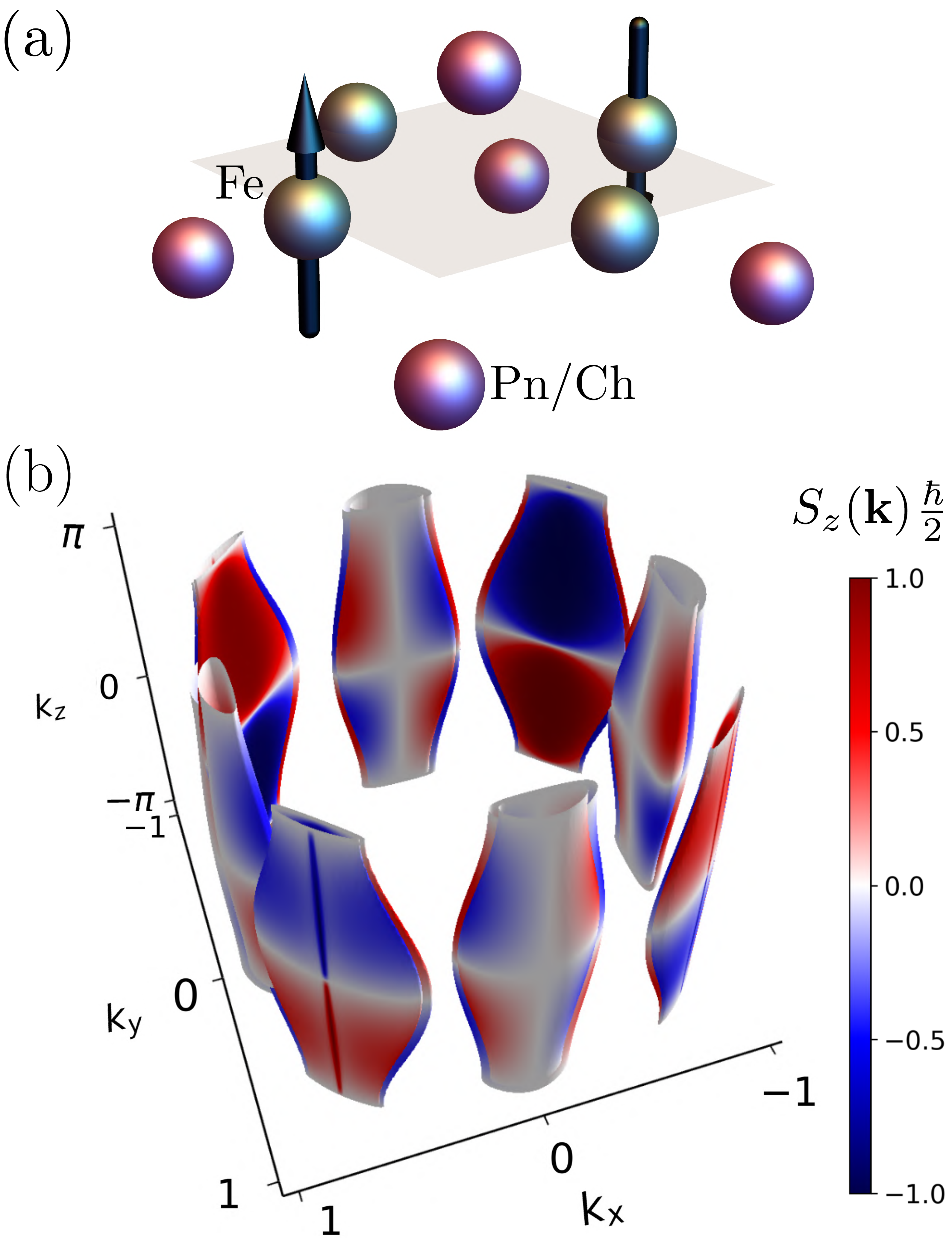}
        \caption{\label{fig:out_of_plane_collinear_Sz}\textbf{Collinear magnetic order.} (a) Illustration of the out-of-plane collinear magnetic order with Fe- and Pn/Ch-atoms included. (b) Fermi surface in the collinear magnetic phase with SOC colored according to $ S_z (\mbf{k})$ which exhibits an $f$-wave form factor. The in-plane components transform as $p$-wave, as in the coplanar case with SOC. Here, $\Delta=60$ meV.}
\end{figure}

The out-of-plane collinear magnetic phase, depicted in Fig.~\ref{fig:out_of_plane_collinear_Sz}(a), also breaks inversion symmetry, and this phase is also realized in the phase diagram of Ref.~\cite{Stadel2022Multiple}. However, no odd-parity magnetism emerges in this case as the collinear structure of the order leaves a residual spin-rotational symmetry which enforces doubly degenerate electronic bands~\cite{Smejkal2022Beyond}. Including SOC breaks this symmetry and splits the bands. In the low-energy model, the magnetic order parameter is
\begin{align}
    \mathcal{H}_{\rm collinear} &= \Delta \sum_{\mathbf{k}}\sum_{\alpha\beta} \Big[ \psi^{\dagger}_{xz,\alpha}(\mathbf{k})\sigma^z_{\alpha\beta}\psi_{xz,\beta}(\mathbf{k}+\mathbf{Q}) \nonumber \\ & + \psi^{\dagger}_{yz,\alpha}(\mathbf{k})\sigma^z_{\alpha\beta}\psi_{yz,\beta}(\mathbf{k}+\mathbf{Q}) \Big] +\text{H.c}\,. \label{eq:out_of_plane_order}
\end{align}

In the presence of SOC the residual spin-rotational symmetry is broken and the out-of-plane momentum-space spin-projection, $S_z (\mathbf{k})$, exhibits an $f$-wave form-factor, as shown in Fig.~\ref{fig:out_of_plane_collinear_Sz}(b). The in-plane components are $p$-wave, as in the coplanar case, and these are not shown. In this case, the Fermi surface splitting is smaller than in the case of the coplanar magnetic phase and it is primarily dependent on the value of the SOC and not the magnetic order parameter $\Delta$. As the spin-splitting is reliant on SOC, this case falls outside the typical classifications of odd-parity magnets~\cite{Hellenes2023P-wave}.

\section{Response and Transport properties}\label{sec:response}

\subsection{Edelstein effect}\label{sec:edelstein}

Odd-parity magnetic systems are potentially relevant for applications in spintronics where an efficient charge-to-spin conversion is essential~\cite{Zutic2004Spintronics}. Indeed, $p$-wave odd-parity magnets feature a large Edelstein effect of a non-relativistic origin (see, e.g., Refs.~\cite{Hellenes2023P-wave,Chakraborty2024Highly,Yu2025Odd-parity}). In contrast, the fact that the spin splitting in $h$-wave odd-parity magnets is antisymmetric under all mirror planes implies that the non-relativistic Edelstein effect vanishes identically in these systems.

\begin{figure}[t]
    \centering
    \includegraphics[width=\linewidth]{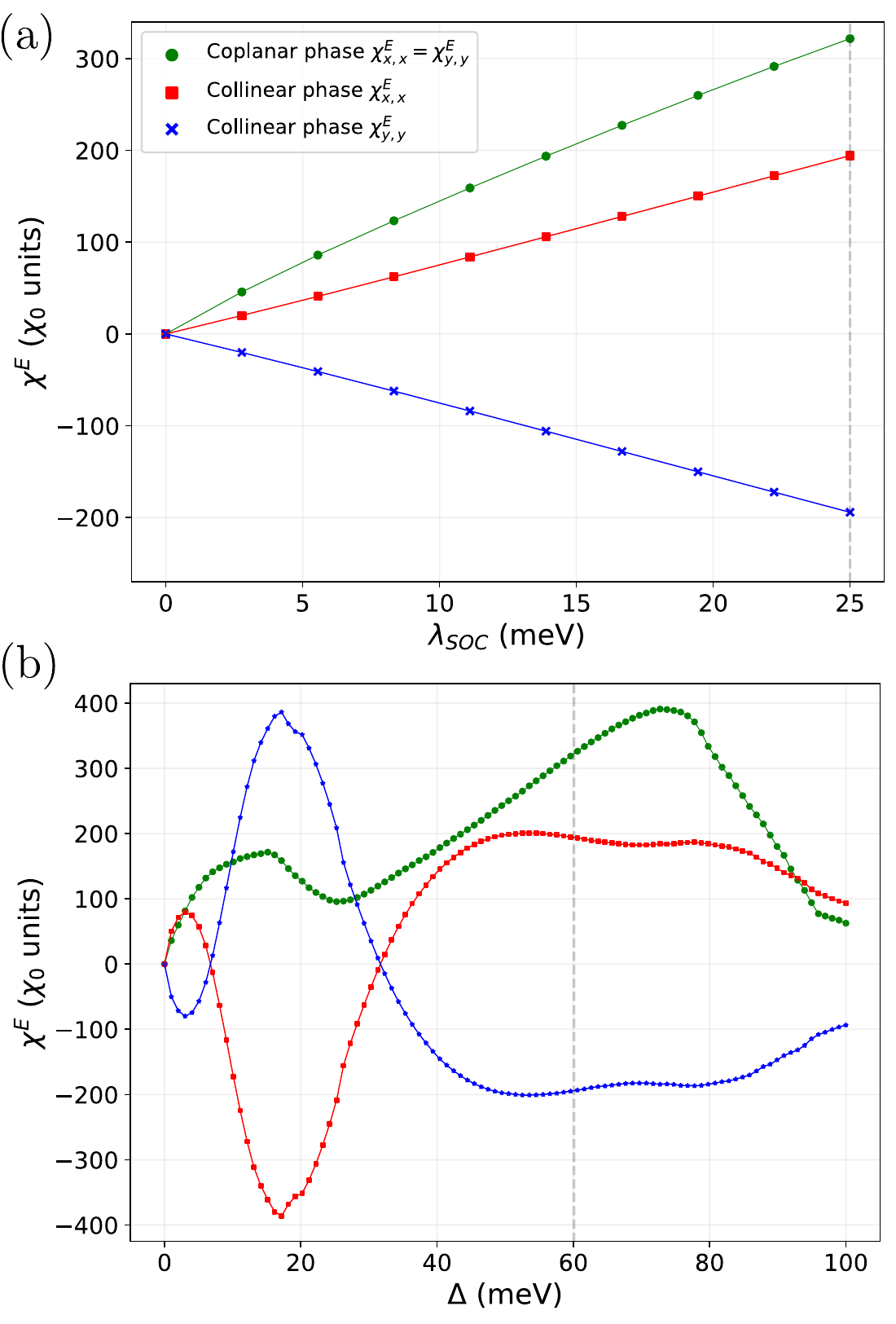}
    \caption{\label{fig:edelstein_delta} \textbf{Edelstein susceptibility}. (a) $\chi^E$ as a function of $\lambda_{\rm SOC}=\lambda_{\rm M}=\lambda_{\Gamma}$ at constant magnetic order parameter, $\Delta=60$ meV. Note that only in-plane components are finite. (b) $\chi^E$ as a function of magnetic order parameter, $\Delta$ at fixed SOC, $\lambda_{\rm M}=\lambda_{\Gamma}=25$ meV. Note that the in-plane spin components refer to the rotated coordinate system defined below Eq.~\eqref{eq:vestigial_SOC}.}
\end{figure}

To see this, we explicitly calculate the Edelstein susceptibility tensor in the presence of coplanar magnetic order, both with and without SOC. The Edelstein susceptibility is the transport coefficient relating an applied electric field to a spin accumulation. It can be obtained from the Boltzmann equation in the relaxation-time approximation by expanding to first order in the electric field yielding the Fermi surface integral
\begin{equation}
    \chi^{E}_{ij} = \chi_0 \sum_{\nu}\int_{FS_\nu}d^2k\,v_i(\mathbf{k})S_j (\mathbf{k})\,. \label{eq:FS_Edelstein_integral}
\end{equation}
Here, $\chi_0=\tau |e| \mu_B A / (2\pi)^2$ with $A$ the area of the unit cell~\cite{Yu2025Odd-parity} and $\tau$ the collision time, which has to be inferred from other experimental probes. $ S_j (\mathbf{k})$ is the momentum-space spin-polarization along the $j$th direction and $v_i(\mathbf{k})$ is the $i$th component of the Fermi velocity. 

The components $\chi^{E}_{zi}$ and $\chi^{E}_{jz}$ are zero due to cancellations caused by the in-plane mirror symmetries $\mathcal{M}_{100}$ and and $\mathcal{M}_{010}$. Here, the components of the Fermi velocity, $v_i$, transform as $p_i$. Consequently, these cancellations occur both without and with weak SOC and in both the coplanar and the out-of-plane collinear phases. In the absence of SOC, the in-plane components of the Edelstein tensor vanish identically because $S_{x,y}(\mathbf{k}) = 0$. When SOC is present, these components become nonzero. In Fig.~\ref{fig:edelstein_delta}(a) we plot the $\chi_{x, x}^{E}$ and $\chi_{y, y}^{E}$ components of the in-plane Edelstein susceptibility tensor for both the coplanar phase and the out-of-plane collinear phase as a function of SOC, for a fixed magnetic order parameter. All other components vanish identically, due to orthogonality of the spin-projection and group velocity form factors [$v_i(\mbf{k})$ transforms as $p_i$]. The Edelstein susceptibility shows a monotonous dependence on the SOC and the effect vanishes when SOC is absent, indicating the relativistic origin of the effect for these systems. In contrast, as shown in Fig.~\ref{fig:edelstein_delta}(b), the dependence on the magnetic order parameter is non-monotonous. We note that, as this is a $\mbf{k}\cdot \mbf{p}$-model, the filling is not constant over the range of $\Delta$ values. Consequently, large values of $\Delta$ will impact the filling and this effect is not captured in Fig.~\ref{fig:edelstein_delta}.

The results presented in Fig.~\ref{fig:edelstein_delta} presents a way of distinguishing the two magnetic orders. This is useful as distinguishing the two magnetic phases typically requires a combination of local probes~\cite{Allred2016Double-Q,Meier2018Hedgehog}. We find that, in the coplanar phase, the in-plane components are identical whereas in the out-of-plane collinear phase, they differ by a sign. This difference can be traced to the composite order parameters $\langle \mbf{M}_1 \times \mbf{M}_2 \rangle \cdot \hat{z}$ and $\langle \mbf{M}_1 \cdot \mbf{M}_2 \rangle$~\cite{Christensen2019Intertwined,Yu2025Odd-parity}, where $\mbf{M}_{1,2}$ denote the two three-component magnetic order parameters. In the coplanar phase, $\langle \mbf{M}_1 \times \mbf{M}_2 \rangle\cdot \hat{z}$ is finite and transforms as $A_{2u}$, thus preserving the $C_4$ rotational symmetry. In contrast, in the collinear phase, $\langle \mbf{M}_1 \cdot \mbf{M}_2 \rangle$ is finite and transforms as $B_{2u}$ which is the origin of the sign difference in Fig.~\ref{fig:edelstein_delta}.

\subsection{Intrinsic non-linear Hall effect}\label{sec:non-linear-Hall}


In crystals that lack inversion symmetry but preserve time-reversal symmetry, the non-linear Hall effect can arise \cite{Sodemann2015Quantum}. The current response to an applied electric field can be expanded as
\begin{align}
    j_a &= \sigma^{(1)}_{ab}\,E_b + \sigma^{(2)}_{abc} E_b E_c \,.
\end{align}
The intrinsic contribution to the second-order conductivity $\sigma^{(2)}_{abc}$ depends on the Berry curvature weighted by derivatives of the Fermi distribution function. This motivates the definition of a quantity known as the Berry curvature dipole~\cite{Sodemann2015Quantum,Du2021Nonlinear},
\begin{align}
    D_{ab} &= \sum_n \int_{\text{BZ}} \frac{d^3k}{(2\pi)^3} \partial_{k_a} f_0(\varepsilon_n(\mathbf{k}))\Omega^b_n(\mathbf{k})\,,
    \label{eq:BCD_derivative_on_fermi_function}
\end{align}
such that
\begin{align}
    \sigma^{(2)}_{abc} 
    &= \frac{e^3\tau}{2\hbar^2(1+i\omega\tau)} \Big( \epsilon_{adc} D_{bd} + \epsilon_{adb} D_{cd} \Big)\,.
    \label{eq:non_linear_cond_tensor}
\end{align}
On integrating \eqref{eq:BCD_derivative_on_fermi_function} by parts, the derivative can instead be applied on the Berry curvature, giving
\begin{equation}
D_{ab} = -\sum_n\int_{\text{BZ}} \frac{d^3k}{(2\pi)^3}  f_0(\varepsilon_n(\mathbf{k})) \partial_{k_a} \Omega^b_n(\mathbf{k}). 
\end{equation}
The integral above is a measure of the first real-space moment of the Berry curvature, hence the name Berry curvature dipole. Here, $\Omega^b_n(\mathbf{k})$ is the Berry curvature pseudovector
\begin{equation}
    \Omega_{c,n} = \frac{1}{2}\varepsilon_{abc}\Omega_n^{ab}\,,
\end{equation}
with the Berry curvature for band $n$ given by
\begin{equation}
    \Omega^{ab}_{n}(\mathbf{k}) = -2\,\mathrm{Im} \sum_{m \neq n} \frac{\langle u_n | \partial_{k_a} H | u_m \rangle \langle u_m | \partial_{k_b} H | u_n \rangle}{(\varepsilon_n - \varepsilon_m)^2}\,.
\end{equation}

To lowest order in temperature, Eq.~\eqref{eq:BCD_derivative_on_fermi_function} reduces to a Fermi surface integral given by
\begin{equation}
    D_{ab} = 
    - \sum_n \int_{\text{FS}_n} \frac{dS}{(2\pi)^3}  \frac{v_{n,a}(\mathbf{k})}{|\mathbf{v}_n(\mathbf{k})|} \Omega^b_n(\mathbf{k})\,. \label{eq:FS_integral_BCD}
\end{equation}
The $z$-component of the Berry curvature pseudovector transforms as $h$-wave in the coplanar phase and $f$-wave in the collinear phase, which implies that its Fermi surface integral vanishes in systems with in-plane mirror symmetry. Consequently, the out-of-plane components of the Berry curvature dipole vanish. This is confirmed by the explicit calculation of the components of the Berry curvature pseudovector for the two bands crossing the Fermi level. $\Omega_z$ exhibits the expected behavior, while $\Omega_y$ and $\Omega_x$ transform as $p_x$ and $p_y$, respectively. As the group velocity component $v_i(\mbf{k})$ transforms as $p_i$, this implies that the only non-zero components of the Berry curvature dipole are $D_{xy}$ and $D_{yx}$.

Keeping in mind that the non-linear conductivity tensor, Eq.~\eqref{eq:non_linear_cond_tensor} is symmetric in its last two indices, we find that the finite components of the non-linear conductivity tensor are $\sigma^{(2)}_{zxx}$, $\sigma^{(2)}_{zyy}$ and $\sigma^{(2)}_{xzx}$, $\sigma^{(2)}_{yzy}$ with the former two being twice as large in magnitude as the latter two.

\section{First-principles results}\label{sec:dft}

\begin{figure}
    \centering
    \includegraphics[width=\columnwidth]{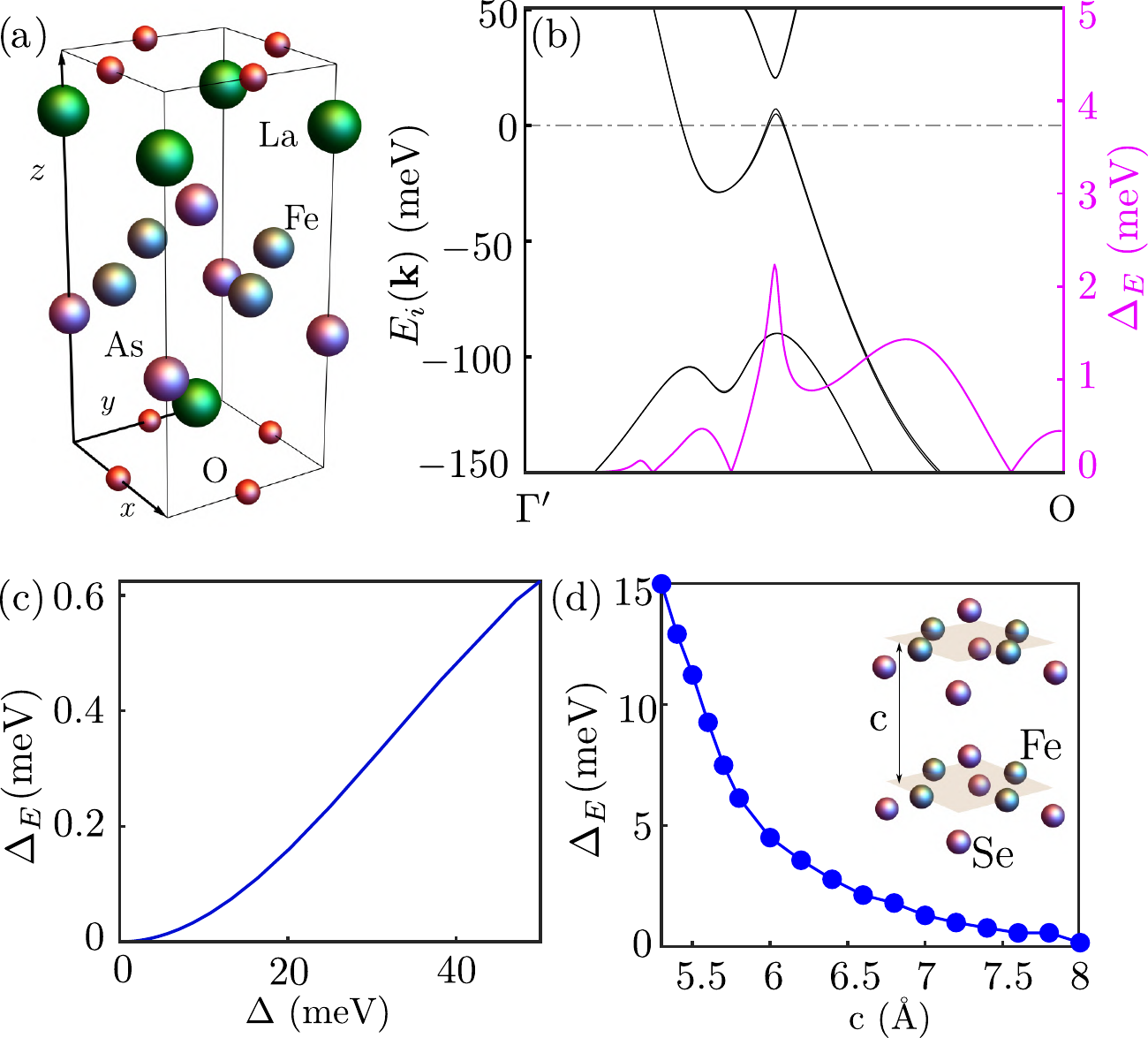}
    \caption{\label{fig:dft_results} \textbf{First-principles results.} (a) Crystal structure of LaFeAsO. (b) Electronic structure of LaFeAsO obtained from DFT along the cut $\Gamma'=(0,0,0.3)\pi$ to $\mathrm{O}=(0.28\sqrt{2},0.42\sqrt{2},0.3)\pi$ with coplanar magnetic order imposed. The electronic bands are plotted in black while the spin splitting of the bands nearest the Fermi level is shown in magenta. (c) Spin splitting as a function of the magnetic order parameter for LaFeAsO for a $\mbf{k}$-point where the Fermi surface intersects the $\Gamma'-\mathrm{O}$-line. (d) Maximum spin splitting along the $\Gamma'-\mathrm{O}$-line for FeSe in the coplanar phase as a function of the separation between neighboring FeSe-layers for $\Delta=100$ meV.}
\end{figure}

We perform \emph{ab initio} calculations for nonmagnetic LaFeAsO, shown in Fig.~\ref{fig:dft_results}(a), using the full-potential local-orbital (FPLO) code~\cite{Koepernik1999Full-potential} in the non-relativistic setting and the LSDA approximation. We use $a=b=4.026492 \mathrm\AA$, $c=9.046755\mathrm \AA$ in space group $P4/nmm$ (\#129). Here, La is at Wyckoff site $2c$ with internal coordinates $(1/4,1/4,0.1416)$, Fe is at $2b$ with $(1/4,1/4,1/2)$, As is at $2c$ with $(1/4,1/4,0.6508)$ and O is at $2a$ with $(1/4, 3/4,0)$. From the resulting electronic structure, we construct a tight-binding model based on the 10 Fe $d$-orbitals (five from each Fe site) labeled $\alpha$ and $\beta$ defined in a local coordinate system with axes along the Fe-Fe nearest-neighbor directions and cut the hopping parameters $t_\delta^{\alpha\beta}$ at a distance of $|\mathbf R_\delta| =40 \mathrm\AA$. The tight binding model then defines a Bloch Hamiltonian (matrix) as usual,
\begin{align}
 H_{\mathbf k}^{\alpha \beta}=\sum_\delta t_\delta^{\alpha\beta}e^{-i\mathbf k\cdot \mathbf R_\delta}\,.
\end{align}
We then introduce a coplanar magnetic order through the term
\newcommand\identity{1\kern-0.25em\text{l}}
\begin{align}
 \widehat\Delta= \Delta \Bigl[\begin{pmatrix} 1 & 0\\ 0 & 0 \end{pmatrix}\otimes \identity_5 \otimes \sigma_x +\begin{pmatrix} 0 & 0\\ 0 & 1 \end{pmatrix}\otimes \identity_5 \otimes \sigma_y \Bigr]\,,
\end{align}
where the matrices refer to the two Fe sites, the five Fe $d$-orbitals, and electronic spin, respectively. The full Bloch Hamiltonian matrix is then
\begin{align}
 H^{\Delta}_\mathbf k=\begin{pmatrix} H_\mathbf k\otimes \sigma_0 & \widehat\Delta \\ \widehat\Delta^\dagger & H_{\mathbf k+\mathbf Q}\otimes \sigma_0 \end{pmatrix}
\end{align}
which describes the odd parity magnet assuming a constant (in orbital space) magnetic moment. Diagonalizing the $40\times 40$ matrix, we obtain the band energies in the odd parity magnetic phase. As expected, the splitting exhibits an $h$-wave structure in momentum space. We quantify the spin splitting $\Delta_E$ as the splitting at a point on the Fermi surface along the path $\Gamma'$ to $\mathrm O$ between bands $\#23$ and $\#24$  which cross the Fermi level and are therefore relevant for the spin splitting at low energies. We choose $\Gamma'=(0,0,0.3)\pi $ and $\mathrm O=(0.28\sqrt 2, 0.42\sqrt 2, 0.3 )\pi$. Fig.~\ref{fig:dft_results}(b) shows the electronic structure along this cut in black, and the spin splitting, $\Delta_E$ in magenta.

Figure~\ref{fig:dft_results}(c) shows the spin-splitting at a specific $\mbf{k}$-point on the Fermi surface as a function of the magnitude of the magnetic order parameter, $\Delta$. Consistent with Eq.~\eqref{eq:spin_splitting}, the behavior is quadratic for small values of $\Delta$, while it becomes linear at large $\Delta$. In the regime where the dependence is quadratic, the magnetic order does not severely distort the bands and the method of imposing a magnetic order on nonmagnetic DFT results remains trustworthy. From Fig.~\ref{fig:dft_results}(c) we find $\Delta_E \approx \alpha \Delta^2$ with $\alpha = 0.44\cdot 10^{-3} \, \mathrm{meV}^{-1}$ in this regime. Relating the order parameter to the magnetic moment requires estimating the magnitude of the interactions. With a conservative estimate of the on-site repulsion $U \approx 1 \, \mathrm{eV}$~\cite{Miyake2010Comparison}, we find that the experimentally observed $0.2\,\mu_B$~\cite{Stadel2022Multiple} yields a splitting of a few meV. 

According to Eq.~\eqref{eq:spin_splitting}, the spin splitting decreases as the anisotropic hopping, $g_1$, is decreased. To test this prediction with DFT, we consider the simpler compound FeSe and vary the separation between the FeSe-layers. For the construction of the tight-binding models for FeSe, we use $a=b=3.7707 \mathrm\AA$, $c=5.521\mathrm \AA$ in space group $P4/nmm$ (\#129). Fe is at Wyckoff site 2b with internal coordinates $(1/4,1/4,1/2)$ while Se is at Wyckoff site 2c with $(1/4,1/4,0.26668)$. For the spacing calculations, the buckling of the Se atoms is kept fixed, while the distance between the FeSe layers is varied. 
The result, shown in Fig.~\ref{fig:dft_results}(d), shows an approximate exponential dependence of the spin-splitting on the layer separation. As the hopping parameters originate from the overlap of exponentially-decaying Wannier functions, this result is consistent with Eq.~\eqref{eq:spin_splitting}. Note that we use FeSe for this comparison only because it is structurally simpler than LaFeAsO. To date, no coplanar magnetic order has been observed in FeSe.

\section{Discussion and conclusions}\label{sec:conclusions}

We have focused on Fe-based superconductors in the space group $P4/nmm$ which exhibits odd-parity magnetism in the coplanar phase due to the nonsymmorphic nature of the space group~\cite{Yu2025Odd-parity}. Another possibility is odd-parity magnetism induced by a helical magnetic order along the $c$-axis with an ordering vector $\mbf{Q}=(0,0,q)$ where $q<\pi/c$. While such a phase has not yet been found in the $P4/nmm$ Fe-based superconductors, it is realized in, e.g., EuRbFe$_4$As$_4$~\cite{Iida2019Coexisting} which is in space group $P4/mmm$ ($\# 123$), where the Eu moments (at Wyckoff site $1a$) exhibit an out-of-plane helical order. Such helical ordering is also observed in a family of non-superconducting Co-based arsenides, including Sr(Co$_{1-x}$Ni$_x$)$_2$As$_2$~\cite{Wilde2019Helical,Nedic2023Competing} and Ca$_{1-x}$Sr$_x$Co$_2$As$_2$~\cite{Li2019Antiferromagnetic} in space group $I4/mmm$ ($\# 139$), where the Co atoms (at Wyckoff site $2b$) carry the magnetic moment.

We have demonstrated that Fe-based superconductors are platforms for $h$-wave odd-parity magnetism, using both a generic symmetry-constrained low-energy model and material-specific DFT calculations. Within the low-energy model, we obtained an analytical expression for the odd-parity spin-splitting, Eq.~\eqref{eq:spin_splitting}, which highlights the importance of the band maxima and minima at $\Gamma$ and M ($\varepsilon_{\Gamma}$ and $\varepsilon_1$), and of the inter-layer coupling through $g_1$. Comparison with DFT results validates the trends predicted by Eq.~\eqref{eq:spin_splitting}. In the absence of SOC, the Edelstein effect vanishes identically in these systems due to the $h$-wave spin-texture form-factor being odd under all mirror symmetries. A finite SOC yields a finite in-plane Edelstein effect which, as shown in Sec.~\ref{sec:edelstein} is different between the coplanar and collinear phases, providing a possible method of distinguishing the two phases using transport measurements.

Our work establishes Fe-based superconductors with coplanar magnetic order as odd-parity magnets. This provides a realization of systems where odd-parity magnetism coexists with unconventional superconductivity. Such odd-parity spin textures are known to derive a distinct superconducting phenomenology~\cite{Smidman2017Review} -- including singlet-triplet mixing and spin-locked Cooper pairs~\cite{Sun2025Ising,Khodas2026Nonrelativistic-Ising} -- while the presence of a non-zero Edelstein effect implies superconducting magnetoelectric responses such as the diode effect~\cite{Shaffer2025Diode}. This is especially relevant as coplanar magnetic order seemingly coexists with unconventional superconductivity in LaFeAs$_{1-x}$P$_x$O~\cite{Stadel2022Multiple}.

\begin{acknowledgments}
We are grateful to Tatsuya Shishidou and Michael Weinert for inspiring discussions. 
R.D. and M.H.C. are supported by ERC grant project 101164202 -- SuperSOC. Funded by the European Union. Views and opinions expressed are however those of the authors only and do not necessarily reflect those of the European Union or the European Research Council Executive Agency. Neither the European Union nor the granting authority can be held responsible for them. A.K. acknowledges support by the Danish National Committee for Research Infrastructure (NUFI) through the ESS-Lighthouse Q-MAT.  B.M.A. acknowledges support from the Independent Research Fund Denmark Grant No. 5241-00007B. D.F.A. was supported by the Department of Energy, Office of Basic Energy Science, Division of Materials Sciences and Engineering under Award No DE-SC0021971.
\end{acknowledgments}

\section{Data availability}

The code used to analyze the low-energy model is openly available~\cite{low-energy_repository} as are the input files for the FPLO calculations for LaFeAsO and FeSe~\cite{dft_repository}.

\appendix

\section{Spin splitting in the coplanar phase}\label{app:perturbation_theory}

Here we provide details of the derivation of the expression for the spin-splitting in Eq.~\eqref{eq:spin_splitting} as well as the crossover from quadratic to linear dependence of the spin-splitting on $\Delta$.
In the low-energy model, the coplanar order reads
\begin{equation}
    \mathcal{H}_{\rm mag} = \sum_{\mathbf{k}}\sum_{\alpha\beta} \Psi^{\dagger}_{\Gamma,\alpha}(\mbf{k}) h_{\rm mag}^{\alpha\beta} \Psi_{M_1,\beta}(\mbf{k}+\mbf{Q}_M) + \text{H.c.}\,,
\end{equation}
with
\begin{equation}
    h_{\rm mag}^{\alpha\beta} = \Delta\begin{pmatrix}
        0 & \sigma^x_{\alpha\beta} \\
        -\sigma^y_{\alpha\beta} & 0
    \end{pmatrix}\,,
\end{equation}
where the minus in front of $\sigma^y_{\alpha\beta}$ is due to the conventional minus in the $E_g$ doublet, Eq.~\eqref{eq:Gamma_doublet}. The full system is described by a $12 \times 12$ Hamiltonian (six orbital states and two spin states). In the absence of SOC, the $12 \times 12$ system can be decoupled into two $6 \times 6$ systems:
\begin{equation}
    H_{12\times12}(\mbf{k}) = H_{\Gamma_{\uparrow}M_{1, 3\, \downarrow}}(\mbf{k}) \oplus  H_{\Gamma_{\downarrow}M_{1, 3\, \uparrow}}(\mbf{k})\,,
\end{equation}
where
\begin{align}
    & H_{\Gamma_{\uparrow}M_{1, 3\, \downarrow}}(\mbf{k}) = \nonumber \\ & \begin{pmatrix}
        h_{\Gamma}^{\uparrow\uparrow}(\mbf{k}) & h^{\uparrow\downarrow}_{\rm mag} & 0 \\
        \left( h^{\uparrow\downarrow}_{\rm mag}\right)^{\dagger} & h^{\downarrow\downarrow}_{M_1}(\mbf{k}+\mbf{Q}_M) & h^{\downarrow\downarrow}_{M_1 M_3}(\mbf{k}+\mbf{Q}_M) \\
        0 & \left( h^{\downarrow\downarrow}_{M_1 M_3}(\mbf{k}+\mbf{Q}_M) \right)^{\dagger} & h^{\downarrow\downarrow}_{M_3}(\mbf{k}+\mbf{Q}_M)
    \end{pmatrix}\,,\label{eq:6x6_matrix}
\end{align}
and similarly for $H_{\Gamma_{\downarrow}M_{1, 3\, \uparrow}}(\mbf{k})$. 

We now treat the three $2\times2$ matrices on the diagonal in Eq.~\eqref{eq:6x6_matrix} as the unperturbed Hamiltonian and the off-diagonal blocks $h^{\alpha\beta}_{\rm mag}$ and $h^{\alpha\beta}_{M_1 M_3}(\mbf{k}+\mbf{Q}_M)$ as perturbations. The eigenvalues and eigenstates of the unperturbed Hamiltonian can be obtained in a straightforward manner and we note that the eigenvalues of the unperturbed $H_{\Gamma_{\uparrow}M_{1, 3\, \downarrow}}(\mbf{k})$ and $H_{\Gamma_{\downarrow}M_{1, 3\, \uparrow}}(\mbf{k})$ are identical. The six different unperturbed eigenvalues are
\begin{align}
    \lambda^{\pm}_{\Gamma(0)} &= \varepsilon_{\Gamma} + \frac{k_x^2 + k_y^2}{2m_{\Gamma}} + t_{\Gamma,z}(1 - \cos k_z) \nonumber \\ & \pm \sqrt{b^2 k_x^2 k_y^2 + c^2 (k_x^2 - k_y^2)^2}\,, \\
    \lambda^{\pm}_{M_1(0)} &= \varepsilon_1 + \frac{k_x^2+k_y^2}{2m_1} + t_{M_1,z}(1-\cos k_z a) \nonumber \\ & \pm \sqrt{a_1^2 k_x^2 k_y^2 + g_1^2 k_x^2 k_y^2 \sin^2 k_z} \\
    \lambda^{\pm}_{M_3(0)} &= \varepsilon_3 + \frac{k_x^2+k_y^2}{2m_3} + t_{M_3,z}(1-\cos k_z a)  \pm a_3 k_x k_y\,.
\end{align}
A finite magnetic order, $\Delta$, will split the eigenvalues. Consequently, the spin-splitting is obtained from the difference between eigenvalues of the perturbed Hamiltonians $H_{\Gamma_{\uparrow}M_{1, 3\, \downarrow}}(\mbf{k})$ and $H_{\Gamma_{\downarrow}M_{1, 3\, \uparrow}}(\mbf{k})$.

\begin{figure}[t]
    \centering
    \includegraphics[width=\columnwidth]{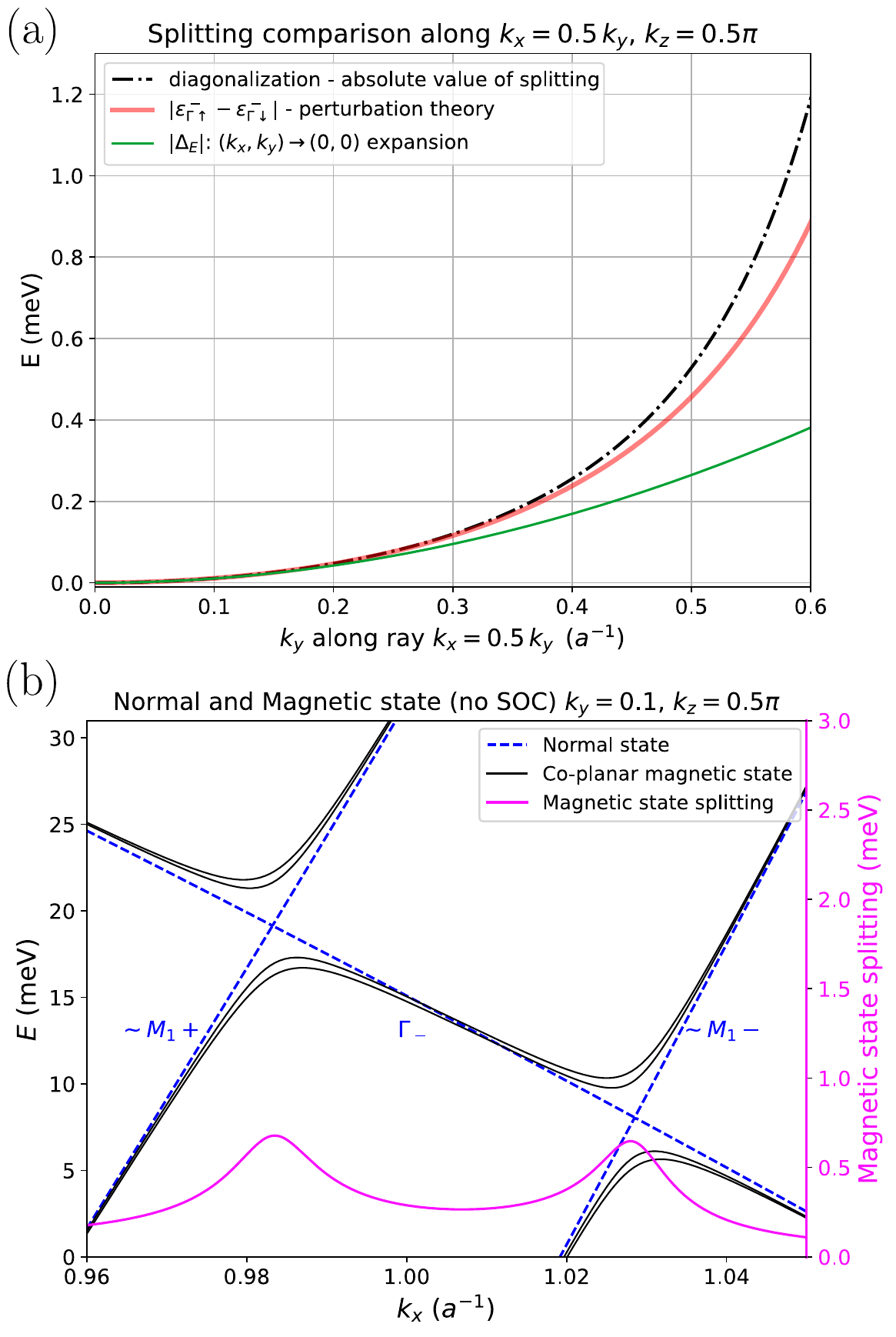}
    \caption{\label{fig:spin_splitting} \textbf{Validation of approximate expression for the spin-splitting} (a) Comparison of the spin splitting obtained from the full diagonalization of the Hamiltonian, the second-order perturbation theory of Eq.~\eqref{eq:spin_splitting_no_k_exp}, and the small $k$-expansion in Eq.~\eqref{eq:spin_splitting_supp}. Here, $\Delta=60$ meV. (b) Plot of the spin splitting at fixed $k_y$ and $k_z$ as a function of $k_x$. The spin splitting peaks at normal state band crossing. Here, we use $\Delta=5$ meV to make the gap opening in the coplanar phase apparent.}
\end{figure}

At first order in perturbation theory there is no spin splitting resulting from $\Delta$ and we therefore apply second-order perturbation theory. For the splitting between the bands originating from the states at $\Gamma$ we find
\begin{equation}
    \lambda^{\pm}_{\Gamma\uparrow}-\lambda^{\pm}_{\Gamma\downarrow} = |\Delta|^2\frac{4\alpha_{\pm}}{1+ \alpha_{\pm}^2} \sum_{s\in\{+,-\}}\frac{\beta_s}{(1+\beta_s^2)(\lambda_{\Gamma(0)}^{{\pm}} - \lambda_{M_1(0)}^{{s}})} \label{eq:spin_splitting_no_k_exp}
\end{equation}
where we have defined 
\begin{align}
    \alpha_{\pm} &= \frac{c(k_x^2 - k_y^2)}{b k_x k_y \mp \sqrt{b^2 k_x^2 k_y^2 + c^2(k_x^2 - k_y^2)^2}} \\
    \beta_{\pm} &= \frac{g_1 k_x k_y \sin k_z}{a_1 k_x k_y \mp \sqrt{a_1^2 k_x^2 k_y^2 +g_1^2 k_x^2 k_y^2 \sin^2 k_z}}
\end{align}
Focusing on $\lambda^{-}_{\Gamma\uparrow}-\lambda^{-}_{\Gamma\downarrow}$, which corresponds to the bands crossing the Fermi level, we expand in powers of $k$ and find
\begin{align}
    \Delta_E &\approx  |\Delta|^2\frac{ 2 c g_1 a_1 f(k)\cdot{(k_x^2 - k_y^2) k_x k_y \sin k_z} }{\left(\lambda^{-}_{\Gamma(0)}- \lambda^{+}_{M_1(0)}\right)\left(\lambda^{-}_{\Gamma(0)}- \lambda^{-}_{M_1(0)}\right)}\label{eq:spin_splitting_intermediate_step}\\&\sim |\Delta|^2\frac{ 2 c g_1 a_1}{(\varepsilon_\Gamma - \varepsilon_1)^2} f(k)\cdot{(k_x^2 - k_y^2) k_x k_y \sin k_z }\,, \label{eq:spin_splitting_supp}
\end{align}
where
\begin{align}
    f(k)^{-1} &= (k_x^2 + k_y^2) \sqrt{\left(\frac{b}{2}\,\sin{2\varphi}\right)^2 + \left(c\,\cos{2\varphi}\right)^2} \nonumber \\ & \times \sqrt{a_1^2+g_1^2\sin^2k_z}\,,
    \label{eq:f_k_function}
\end{align}
with $\varphi$ the polar angle between $k_x$ and $k_y$. In the above expansion we have furthermore assumed $t_{M_1,z} \approx t_{M_3,z}$ to simplify the expressions. The expression in Eq.~\eqref{eq:spin_splitting_supp} clearly exhibits an $h$-wave form factor. In Fig.~\ref{fig:spin_splitting}(a) we show a comparison between the exact splitting obtained from diagonalizing the Hamiltonian numerically, the expression in Eq.~\eqref{eq:spin_splitting_no_k_exp} for $\lambda^{-}_{\Gamma\uparrow}-\lambda^{-}_{\Gamma\downarrow}$, and the small $k$-expansion in Eq.~\eqref{eq:spin_splitting_supp}, $\Delta_E$.

At the unperturbed band crossings, the denominator in Eq.~\eqref{eq:spin_splitting_intermediate_step} vanishes, causing the splitting to formally diverge which is an indication that non-degenerate perturbation theory breaks down at these points. While this approximation cannot provide a reliable quantitative prediction for the magnetic-state spin-splitting exactly at the unperturbed crossings, it nevertheless captures a key qualitative feature: the largest spin splitting in the magnetic state occurs in the vicinity of the $\mbf{k}$-points where the normal-state bands cross, as shown in Fig.~\ref{fig:spin_splitting}(b).


From DFT calculations and diagonalizing the complete $\mbf{k}\cdot \mbf{p}$ model, we observe that the splitting scales linearly with $|\Delta|$ for large values of $\Delta$, while for small values, it follows a quadratic dependence. This trend can be understood within a simple two-band model, where two energy levels $\varepsilon_1$ and $\varepsilon_2$ are coupled via an amplitude $t$. The exact expression for the splitting in this model is $\delta\epsilon = \sqrt{(\varepsilon_1 - \varepsilon_2)^2 + 4|t|^2}$, which, when expanded, yields $\delta\epsilon \sim (\varepsilon_1 - \varepsilon_2) + \frac{2|t|^2}{\varepsilon_1 - \varepsilon_2}$. Comparing this with the structure of the splitting in Eq.~\eqref{eq:spin_splitting_supp}, we propose
\begin{align}
    \Delta_E \sim & \left(\sqrt{(\varepsilon_\Gamma - \varepsilon_1)^2 + 4|\Delta|^2}-|\varepsilon_\Gamma - \varepsilon_1|\right) \frac{cg_1a_1 f(k)}{|\varepsilon_\Gamma - \varepsilon_1|} \nonumber \\ & \times (k_x^2 - k_y^2)\,k_xk_y\,\sin k_z\,. 
\end{align}
For small $\Delta$, this reduces to Eq.~\eqref{eq:spin_splitting_supp} with a parabolic dependence on $|\Delta|^2$, whereas for $\Delta \gg \frac{|\varepsilon_1 - \varepsilon_\Gamma|}{2}$, the splitting becomes linear in $|\Delta|$. Thus, the crossover from parabolic to linear behavior occurs at an energy scale approximately given by $\sim \frac{|\varepsilon_1 - \varepsilon_\Gamma|}{2}$.

\bibliography{uncon_bib}

\end{document}